\documentclass[preprint2]{aastex}

\shorttitle{Nonextensive DM and gas density profiles}
\shortauthors{Leubner}

\begin{document}


\title{Nonextensive theory of dark matter and gas density profiles}


%\author{M. P. Leubner\altaffilmark{1}}
\author{M. P. Leubner}
\affil{Institute of Astrophysics, University of Innsbruck,
    A-6020 Innsbruck, Austria}
\email{manfred.leubner@uibk.ac.at}


%\altaffiltext{1}{Institute of Astrophysics, University of Innsbruck, Austria}


\begin{abstract}
Pronounced core-halo patterns of dark matter and gas density
profiles, observed in relaxed galaxies and clusters, were hitherto
fitted by empirical power-laws. On the other hand,
similar features are well known from astrophysical plasma environments,
subject to long-range interactions, modeled in the context of
nonextensive entropy generalization. We link nonextensive statistics
to the problem of density distributions in large-scale structures
and provide fundamentally derived density profiles, representing accurately
the characteristics of both, dark matter and hot plasma distributions, as
observed or generated in simulations. The bifurcation of the density
distribution into a kinetic dark matter and thermodynamic gas branch 
turns out as natural consequence of the theory and is controlled by a
single parameter kappa, measuring physically the degree of coupling
within the system. Consequently, it is proposed to favor nonextensive
distributions, derived from the fundamental physical context of entropy
generalization and accounting for nonlocality and long-range interactions
in gravitationally coupled systems, when modeling observed density
profiles of astrophysical structures.  
\end{abstract}


\keywords{cosmology:theory---dark matter---galaxies: halos---galaxies:structure---plasmas}


\section{Introduction}

The analysis of dark matter (DM) and gas density distributions in
galaxies and clusters is presently based on a variety of phenomenological
models. Following \citet{King62} relaxed DM halo
density profiles are well fitted by empirical power laws \citep{Burkert95,Salucci00}
and from N-body simulations by $\rho_{DM} \sim (r/r_{s})^{-1} (1+r/r_{s})^{-2}$
(NFW), where $r_s$ constitutes a scaling radius, chosen to join the
asymptotic $r$-dependence \citep{Navarro96,Navarro97}. Subsequently a number
of modifications were proposed \citep{Fukushige97,Moore98,Moore99} along
with criticism of the 'universality' of the NFW-profile \citep{Jing00,Borriello01}.
Comparing the density profiles of DM haloes in cold dark matter (CDM)
N-body simulations the functional dependence 
$\rho_{DM} \sim (r/r_{s})^{-\alpha} (1+r/r_{s})^{-(3-\alpha)}$
\citep{Zhao96} was found to provide good fits to all haloes,
from dwarf galaxies to clusters at any redshift \citep{Ricotti03},
where $\alpha$ is related to the spectral index of the
initial power spectrum of density perturbations. Recently a 
universal density profile for dark and luminous matter was suggested
\citep{Merritt05}. Physically, we regard
the DM halo as a self-gravitating collisionless system of weakly
interacting particles in dynamical equilibrium, scale invariant from
galactic to cluster scales \citep{Burkert00,Firmani00,Spergel00}.  

On the other hand, the phenomenological $\beta-$model
$\rho_{gas} \sim (1+r^2/r_{c}^2)^{-3/2\beta}$ \citep{Cavaliere76}, where
$r_c$ is the core radius, or the double $\beta$-model, a convolution
of two $\beta$-models with the aim of resolving the $\beta-$discrepancy
\citep{Bahcall94},
provide reasonable representations of the hot gas density distribution
in galaxies and clusters \citep{Xue00,Ota04}. Physically, $\beta$ corresponds to
the ratio of kinetic DM and thermal gas energy, assuming values 
$\sim$ $2/3$.

Since any astrophysical system is subject to long-range gravitational
or electromagnetic interactions, the
present situation motivates to introduce nonextensive statistics as
theoretical basis for both, DM and hot plasma density profiles, utilized
successfully to understand observed core-halo structures in astrophysical
plasmas \citep{Leubner04,Leubner05}. In this situation a
single parameter $\kappa$ characterizes the degree of nonextensivity
or coupling within the system. The corresponding derived power-law
distributions constitute a particular
thermodynamic equilibrium state \citep{Treumann99}, commonly applied
in astrophysical plasma modeling \citep{Leubner01,Leubner02}.

The concept of the Boltzmann-Gibbs-Shannon (BGS) thermo-statistics constitutes a
powerful tool whenever the physical system is extensive, i.e. the entropy is additive.
This situation holds when microscopic interactions and memory are short
ranged and the environment is an Euclidean space-time, a continuous and
differentiable manifold. However, astrophysical systems are generally
subject to spatial or temporal long-range interactions making their behavior
nonextensive. A generalization of the BGS entropy
for statistical equilibrium was introduced from first principles by \citet{Renyi55}
and \citet{Tsallis88}, suitably extending the standard additivity to
nonextensivity. The main theorems of the classical Maxwell-Boltzmann
statistics admit profound generalizations within nonextensive statistics
leading to a variety of physical consequences, see e.g.
\citet{Tsallis95}. Those include a reformulation of the classical
N-body problem \citep{Plastino94} or the development of nonextensive
distributions \citep{Silva98,Almeida01,Andrade02} where the duality
of nonextensive statistics, we will focus on, was recognized \citep{Karlin02}.
Astrophysical applications \citep{Plastino93,Kaniadakis96,Nakamichi02},
provided further manifestation of nonextensivity in nature. For a reformulation
in the context of special relativity see \citet{Kaniadakis02}.

\section{Theory}

The generalized entropy $S(\kappa)$ characterizing systems subject to
long-range interactions and couplings in nonextensive statistics reads
\citep{Tsallis88,Leubner04}

\begin{equation}
S_{\kappa }=\kappa k_{B}({\sum }p_{i}^{1-1/\kappa }-1)
\label{1}
\end{equation}

where $p_{i}$ is the probability of the $i^{th}$ microstate, $k_{B}$ is
Boltzmann's constant and the 'entropic index' $\kappa$ denotes a coupling 
parameter quantifying the degree of nonextensivity, or equivalently
statistical correlations, within the system.
A crucial property of this entropy is the pseudo-additivity for given
subsystems in the sense of factorizability of the microstate probabilities.
The transformation $\kappa = 1/(1-q)$ links the $\kappa$-formalism 
to the Tsallis q-statistics
\citep{Leubner02}. Here $\kappa$ is defined in the interval
$-\infty \leq \kappa \leq \infty $ where $\kappa = \infty$ represents
the extensive limit of statistical independence and recovers the classical
BGS entropy as $S_{B}=-k_{B}{\sum p_{i}}\ln p_{i}$.
Considering two subsystems $A$ and $B$ the nonextensive characteristics
can be illuminated in view of the entropy mixing by
$S_{\kappa}(A+B)=S_{\kappa}(A)+S_{\kappa}(B)+ S_{\kappa}(A)S_{\kappa}(B)/\kappa$,
a relation consistent with Eq. (\ref{1}) where the last term
accounts for the couplings. $\kappa < 0$ leads to an entropy
decrease providing a state of higher order, whereas for $\kappa > 0$
the entropy increases and the system evolves into disorder. Hence,
$\kappa$ can be interpreted as a bifurcation parameter
measuring the two statistical realizations of ordering or disordering
through correlations. A generalization for multiple subsystems is
discussed by \citet{Milovanov00}.

Since entropy and probability distributions reside physically on the same
level the corresponding generalized energy distributions follow as 
$f^{\pm}(v)=A^{\pm}\left[ 1+ v^{2}/(\kappa \sigma^{2})\right] ^{-\kappa}$
\citep{Silva98,Leubner04}. The superscripts $\pm$ correspond to positive
or negative values of $\kappa$, $A^{\pm}$ are proper normalization constants,
and $\sigma$ denotes the velocity dispersion or mean energy of the
distribution characterising their width (variance). Negative values of
$\kappa$ are conveniently introduced by changing the sign at appearance,
which generates a cutoff at $v = \sqrt{\kappa}\sigma$,
see \citet{Leubner04} for details. 

\begin{figure}[t]
%\plotone{f1.eps}
\center
\includegraphics[width=7.5cm]{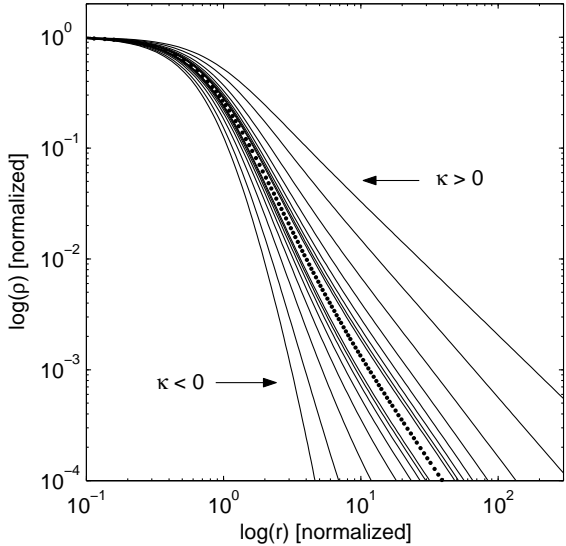}
\caption{Nonextensive density profiles:
for $\kappa=3$ the innermost and outermost curves correspond to the
$\rho^{-}$ and $\rho^{+}$ solutions, respectively. For increasing
$\kappa=4...10$ both sets of curves converge to the
central isothermal sphere solution ($\kappa = \infty$)
indicated by dots, $\rho^{-}$ from inside
and $\rho^{+}$ from outside.
\label{fig1}}
\end{figure}

Upon generalization to a spherical symmetric, self-gravitating and
collisionless N-body system the corresponding steady state phase-space distribution
$f(r,v)$ obeys the Vlasov equation. If the particles
(stellar system, galaxies) itself provide the gravitational potential
and $f(r,v)$ is regarded as the mass distribution then Poisson's equation
$\Delta \Phi =4\pi G\rho$ reads  

\begin{equation}
\Delta \Phi = 4\pi G\int f(\frac{1}{2}v^{2} +\Phi )d^{3}v,
\label{2}
\end{equation}

and represents the fundamental equation governing the equilibrium of the
system, where $f(\frac{1}{2}v^{2}+\Phi)=f(E)$ depends on the energy only.  
Commonly, the relative particles energy 
$E_{r} = -1/2v^2 + \Psi$ is introduced where $\Psi$ is the relative potential 
$\Psi = - \Phi + \Phi_{0}$, chosen to vanish at the systems boundary and
satisfying Poisson's equation as $\Delta \Psi =-4\pi G\rho$.

If $f(E_{r})$ resembles the exponential mass distribution
function defining the structure of an isothermal self-gravitating sphere of
gas, in this case identical to the phase-space density distribution of a
collisionless system of particles,

\begin{equation}
f(E_{r})= \frac{\rho_{0}}{(2\pi \sigma^{2})^{3/2}}  
\exp(-\frac{v^{2}/2-\Psi} {\sigma^{2}})
\label{3}
\end{equation}

then the corresponding density
distribution $\rho = \rho_{0} \exp(\Psi/ \sigma^{2})$ is found after
integrating over all velocities. Combining with Poisson's equation (\ref{2})
the solution governs the structure of the isothermal self-gravitating sphere
\citep{Binney94}. The equilibrium distribution Eq. (\ref{3}) can
be obtained by extremizing the standard BGS entropy with regard to
conservation of mass and energy.

Since Eq. (\ref{3}) applies exclusively to a system of independent
particles we introduce now long-range interactions 
by the generalized entropy functional (\ref{1}). Extremizing
Eq. (\ref{1}) after replacing the entropy function $flnf$ of an uncorrelated
ensemble by the generalized functional
$-\kappa f (1 - f^{1-1/\kappa}$ and applying Lagrange multipliers
\citep{Plastino93} the resulting distribution function reads 

\begin{equation}
f^{\pm}(E_{r})=B^{\pm} \left[ 1+\frac{1}{\kappa}
\frac{v^{2}/2-\Psi} {\sigma^{2}}\right] ^{-\kappa}
\label{4}
\end{equation}

As previously, the superscripts refer to the positive or negative intervals of
the entropic index $\kappa$, accounting for less (+) and higher (-) organized states
and thus reflecting the accompanying entropy increase or decrease, respectively.
If we identify $f^{\pm}(E_{r})$ again as mass
distribution, the $\kappa$-dependent generalized constants $B^{\pm}$ assure
proper normalization and dimension and differ for positive and negative
definite $\kappa$-values as 
$B^{+}=C\Gamma(\kappa)/(\kappa^{3/2}\Gamma(\kappa-3/2))$ and
$B^{-}= C\Gamma(\kappa+5/2)/(\kappa^{3/2}\Gamma(\kappa+1))$, 
where $C=\rho_{0}/(2\pi \sigma^{2})^{3/2}$ and
$\Gamma$ denotes the standard gamma function \citep{Leubner04}. 
The different normalization is caused by the interval corresponding to
negative $\kappa$-values, which generates an energy cutoff in
Eq. (\ref{4}) leading to the constraint
$ v^{2}/2-\Psi \leq \kappa \sigma^{2}$ and restricting also the
integration limits in velocity space. For
$\kappa \rightarrow \infty$ Eq. (\ref{4}) approaches the exponential
distribution function (\ref{3}) defining the density profile of the
isothermal sphere.

After incorporating the sign of $\kappa$ into Eq. (\ref{4}),
we perform separately for positive and negative definite $\kappa$  the
integration of (\ref{4}) over all velocities where $B^{\pm}$
must be used consistently.
The resulting solution provides a modification of the velocity space context
introduced by \citet{Leubner04} for the density evolution of a system in 
gravitational potential as 

\begin{equation}
\rho^{\pm}=\rho_{0}  \left[ 1-\frac{1}{\kappa} \frac{\Psi} {\sigma^{2}}\right] ^{3/2-\kappa}
\label{5}
\end{equation}

In analogy to the corresponding tandem character in velocity space Eq. (\ref{5})
generates in a gravitational potential for finite positive values of $\kappa$
pronounced density tails, whereas for negative $\kappa$-values the solutions
are restricted within the cutoff at $\kappa = \Psi / \sigma^{2}$ and
$\kappa = -\infty$.

\begin{figure}[t]
%\plotone{f2.eps}
\center
\includegraphics[width=7.5cm]{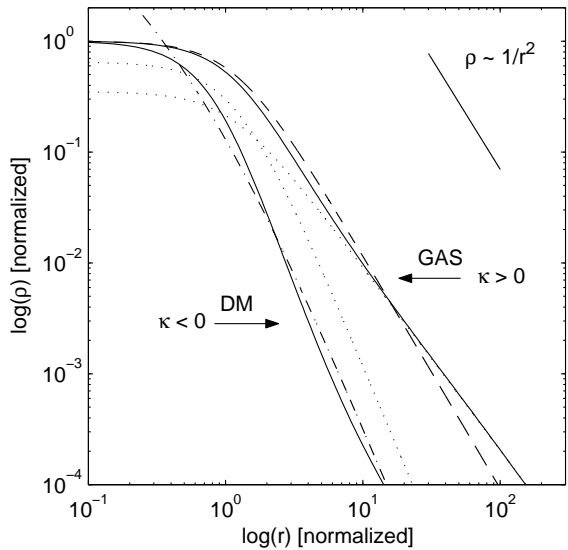}
\caption{Comparison of the DM nonextensive density profile ($\kappa = -7$,
$\sigma=1$, solid) with the NFW-profile ($r_{s}=0.65, \rho_{0}=1.2$,
dashed-dotted). The radial nonextensive gas distribution
($\kappa = 7$) is compared with a single $\beta-$model
($\beta = 0.7$, dashed) and the decomposition of a double $\beta-$model
($\beta_{core} = 0.98$, $\beta_{halo} = 0.55$, dotted). As illustration
a dependence $\rho \simeq 1/r^{2}$ is added.
\label{fig2}}
\end{figure}

The duality of equilibria in nonextensive statistics
is manifest in two families, the nonextensive thermodynamic equilibria
and the equilibria of kinetic equations, where both are related by 
$q^{'} = 2-q$ \citep{Karlin02}. Since
$q=1-1/\kappa$ \citep{Leubner02} we find for the entropic
index $\kappa^{'}=-\kappa$ relating the two families of equilibria,
where with regard to Eq. (\ref{5}) $\kappa > 0$
corresponds to the stationary states of thermodynamics and 
$\kappa < 0$ to kinetic stationary states. The limiting BGS state
for $\kappa = \infty$ is therefore characterized by self-duality.
The nonextensive parameter $\kappa$ finds also a physical interpretation
in terms of the heat capacity of a medium \citep{Almeida01}. A system
with $\kappa > 0$ represents an environment with finite positive heat capacity
and vice versa, for $\kappa < 0$ the heat capacity is negative.
Negative heat capacity is a typical property of self-gravitating
systems, see e.g. \citet{Firmani00}. Moreover, contrary to thermodynamic systems
where the tendency to dis-organization is accompanied by increasing entropy,
self-gravitation tends to result in higher organized structures of decreased
entropy. Consequently, the nonextensive bifurcation of the singular
isothermal sphere solution into two distributions $f^{\pm}(E_{r})$ or
$\rho^{\pm}$, respectively,  requires to identify
the  density profile (\ref{5}) for positive definite $\kappa$ as
the proper distribution of the thermodynamic state of the gas, 
whereas the negative definite counterpart is associated to the
self-gravitating DM distribution. For
$\kappa \rightarrow \infty$ both solutions merge at the
isothermal sphere distribution defined by Eq. (\ref{3}).

After elucidating the tandem character of the nonextensive context
we combine Poisson's equation $\Delta \Psi = -4\pi G\rho$ with
Eq. (\ref{5}) to find a second order nonlinear differential equation
for for the radial density dependence of a spherically symmetric gas and DM
distribution as

%\begin{equation}
\begin{eqnarray}
\frac{d^{2} \rho}{{dr}^{2}}+\frac{2}{r} \frac{d\rho}{dr}-(1-\frac{1}{n})
\frac{1}{\rho}(\frac{d\rho}{dr})^{2}-  \\ \nonumber
\frac{4\pi Gn}{(3/2 - n) \sigma^{2}}\rho^{2}(\frac{\rho}{\rho_{0}})^{-1/n}=0
\label{6}
\end{eqnarray}
%\end{equation}

where $n=3/2-\kappa$ is introduced. n corresponds to the polytropic index
$n=1/(\gamma-1)$ of stellar dynamical systems where $\gamma$ is the adiabatic index. Hence,
the limit $\kappa = \infty$ is governed consistently with $\gamma=1$ by the
equation of state of the isothermal sphere \citep{Binney94} and Eq. (\ref{4})
denotes the nonextensive generalization. Multiplying Eq. (\ref{6}) in the
limit $\kappa \rightarrow \infty$ by $r^{2}/ \rho$ and combining the first
three terms on the left hand side as $d/dr(r^{2} dln\rho/dr)$
recovers the equation defining the conventional isothermal sphere.
Eq. (\ref{6}) is solved numerically by
simulation of the corresponding two first order differential equations.

\begin{figure}[t]
%\plotone{f3.eps}
\center
\includegraphics[width=7.5cm]{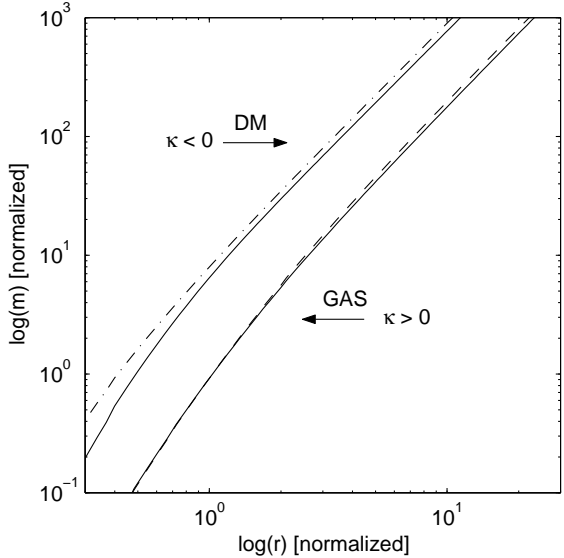}
\caption{Nonextensive integrated mass of DM and gas components
(solid) as compared with the NFW-profile (dashed-dotted) and
the single $\beta-$model (dashed) for the gas component.
Parameters are the same as for Fig. 2.
\label{fig3}}
\end{figure}

\section{Results and discussion}

As natural consequence of nonextensive entropy generalization the
isothermal sphere profile bifurcates into two distribution families
controlled by the sign of $\kappa$ and governed by Eq. (\ref{6}).
The self-gravitating DM component,
a lower entropy state due to gravitational interaction, resides besides
a thermodynamic gas component of increased entropy.
Fig. 1 provides the radial density profile
characteristics and $\kappa$ dependences as computed from Eq. (\ref{6})
for both families, DM distributions located below and the plasma
distributions above the limiting exponential solution.
With increasing entropic index $\kappa$ both braches
merge simultaneously in the isothermal sphere density profile,
since the power-laws (\ref{4}) converge for $\kappa \rightarrow \infty$ 
to the exponential energy distribution (\ref{3}), corresponding to
$n= \infty$ in Eq. (\ref{6}). Physically
finit values of $\kappa$ represent long-range interactions and
correlations within the system, whereas the transition to $\kappa = \infty$
in the central curve defines the extensive limit of statistical independence. 
Since we focus here on the shapes of radial profiles and their physical
foundations, normalizations are conveniently applied. As $r \rightarrow 0$,
$\rho = \rho_{0} = 1$ and the solution meets the physically
required condition $d\rho/dr = 0$ in the origin.

In Fig. 2 we compare one negative (DM) solution to Eq. (\ref{6})
with the NFW model as well as one symmetrically, positive (gas) solution
with a single $\beta-$model. On small scales the theoretical DM density
distribution is characterized consistently with e.g. \citet{Firmani00}
by a shallow core of finite density as $r \rightarrow 0$.
Despite a gradually tail-like structure of the theoretical profile
the halo characteristics are similar to the NFW profile. Changes
in the mean energy or variance $\sigma$ generates a radial shift of the
entire profile, which practically corresponds to variations of the scale radius
$r_{s}$ of the NFW model. As measure of the long-range interactions the
dimensionless second parameter $\kappa$ of the theory controls 
the shape and characteristic mean slope of the profile on intermediate and
large scales, see also Fig. 1. Thus both, shallower or steeper slopes
\citep{Moore98,Moore99}, interpreted in the nonextensive context as
consequence of different coupling strengths, are accessible.
Recently an improved fitting formula, converging to a finite density in the center,
was introduced and compared with the standard NFW-profile as well \citep{Navarro04}.
The accompanying discussion illustrates critically also the potential difficulties 
arising in simulations of the innermost structure of CDM haloes (see also \citet{Trott05}).
On the other hand, high-resolution rotation curve analyses of galaxies are consistent
only with cored haloes \citep{Blok03,Gentile04} supporting the physical
consequences of the nonextensive statistics formulation
from observations. The nonextensive gas density distribution
follows a single $\beta-$model in the core but deviates by a halo tail formation.
This deviation
can be accurately fitted by a double $\beta-$model (a decomposition as in
\citet{Xue00} is shown by the dotted lines), indicating that the
nonextensive theory provides naturally a theoretical context able to solve
the $\beta-$discrepancy. 

Fig. 3 presents the 
radial dependence of the integrated mass of DM and gas components
for symmetric values of $\kappa=\pm 7$ as compared to the NFW 
and $\beta-$model, where on observational grounds a $20 \%$ central gas
density fraction $\rho_{0}$ is introduced for proper visualization.
Consistent with Fig. 2 the NFW integrated mass exceeds the nonextensive
solution slightly, whereas the integrated $\beta-$model is practically
 identical to the generalized entropy approach. 

The dual nature of the nonextensive theory provides a solution
to the problem of DM and gas density distributions of clustered matter
from fundamental physics, where both parameters ($\kappa$, $\sigma$)
admit physical interpretation. The bifurcation of the density distribution
into a kinetic DM and thermodynamic gas branch turns out
as natural consequence of the theory and is controlled by the
entropic index $\kappa$, accounting physically for nonlocality and 
long-range interactions in nonextensive systems. Due to different
correlation properties of clustered structures, particular DM and gas
density profiles might be subject to different values of $\kappa$ and $\sigma$,
regulating details of possibly non-universal, mass dependent profiles. 
The theory was found to reproduce accurately also density profiles
generated by N-body and hydrodynamic simulations, a subject to be
addressed elsewhere. In conclusion, it is proposed to favor the family
of nonextensive distributions, derived from the fundamental physical
context of entropy generalization, over empirical models when
fitting observed density profiles of astrophysical structures.

\end{document}